# Effects of gas diffusion layer thickness on PEM fuel cells with composite foam-rib flow fields

Wei Gao[1], Qifeng Li[1], Kai Sun[1], Rui Chen[1,3], Zhizhao Che[1,2]*, Tianyou Wang[1,2]*

1. State Key Laboratory of Engine, Tianjin University, Tianjin, 300350, China
2. National Industry-Education Platform of Energy Storage, Tianjin University, Tianjin, 300350, China
3. Department of Aeronautical and Automotive Engineering, Loughborough University, Loughborough LE11 3TU, United Kingdom

*Corresponding authors: chezhizhao@tju.edu.cn, wangtianyou@tju.edu.cn

## Abstract

Gas diffusion layers (GDLs) play a crucial role for the performance of proton exchange membrane fuel cells (PEMFCs). The utilization of composite foam-rib flow fields (CFRFFs) can alter the reactant gas transfer pattern, hence improving the efficiency of under-rib reactant gas transfer and water drainage. The impact of the cathode and anode GDL thicknesses ($h_{c,GDL}$ and $h_{a,GDL}$) on the performance of CFRFF design is investigated by three-dimensional multiphase non-isothermal numerical simulation in this study. The results indicate that for the conventional rib flow field (CRFF) design, there is an optimal $h_{c,GDL}$ for optimal cell performance, while for the CFRFF design, as $h_{c,GDL}$ becomes thinner, the cell performance increases, and the trend is dominated by the variation of the oxygen concentration. Under a thin GDL, the rib width of the CRFF design should be as small as possible to minimize concentration polarization loss, while the rib width of the CFRFF design can be slightly larger. Furthermore, by decreasing the thickness of $h_{a,GDL}$ in both the CRFF and CFRFF designs, there is an increase in the dissolved water content in the ionomer of the cathode CL and a subsequent decrease in the Ohmic polarization loss.

# Nomenclature

| | |
|---|---|
| $a$ | water activity (-) |
| $A$ | cell geometric area (cm$^2$) |
| $C_p$ | specific heat (J kg$^{-1}$ K$^{-1}$) |
| $C$ | molar concentration (mol m$^{-3}$) |
| $D$ | mass diffusivity (m$^2$ s$^{-1}$) |
| $EW$ | the equivalent weight of membrane (kg kmol$^{-1}$) |
| $F$ | Faraday's constant (C mol$^{-1}$) |
| $h$ | thickness (mm) |
| $H$ | height (mm), latent heat (J kg$^{-1}$) |
| $I$ | current density (A cm$^{-2}$) |
| $\mathbf{J}$ | the current density vector (A m$^{-2}$) |
| $j$ | volumetric exchange current density (A m$^{-3}$) |
| $K$ | permeability (m$^2$) |
| $k$ | thermal conductivity (W m$^{-1}$ K$^{-1}$) |
| $L$ | length (mm) |
| $M$ | molecular weight (kg mol$^{-1}$) |
| $\dot{m}$ | mass flow rate (kg s$^{-1}$) |
| $n_d$ | electro-osmotic drag coefficient (H$_2$O per H$^+$) |
| $P$ | pressure (Pa) |
| $R$ | universal gas constant (J mol$^{-1}$ K$^{-1}$) |
| $S$ | entropy (J mol$^{-1}$ K$^{-1}$) or source terms |
| $T$ | temperature (K) |
| $\mathbf{u}$ | velocity (m s$^{-1}$) |
| $V$ | electrical potential (V) |
| $W$ | width (mm) |
| $X$ | mole fraction (-) |
| $Y$ | mass fraction (-) |

## Greek letters

| | |
|---|---|
| $\alpha$ | transfer coefficient (-) |
| $\gamma$ | water phase change rate (s$^{-1}$) |

| $\varepsilon$ | porosity (-) |
| --- | --- |
| $\zeta$ | water transfer rate (s$^{-1}$) |
| $\eta$ | overpotential (V) |
| $\theta$ | contact angle (°) |
| $\kappa$ | electrical conductivity (S m$^{-1}$) |
| $\lambda$ | water content in ionomer (-) |
| $\mu$ | dynamic viscosity (kg m$^{-1}$ s$^{-1}$) |
| $\xi$ | stoichiometry ratio (-) |
| $\rho$ | density (kg m$^{-3}$) |
| $\sigma$ | surface tension (N m$^{-1}$) |
| $\phi$ | rib width ratio (-) |
| $\varphi$ | potential (V) or volume fraction (-) |
| $\omega$ | volume fraction of ionomer in the catalyst layer (-) |
| $\Omega$ | computational domain |

## Subscripts

| 0 | intrinsic property or environmental conditions |
| --- | --- |
| a | anode |
| act | active overpotential |
| a,end | anode bipolar plate |
| a,gdl | anode gas diffusion layer |
| bp | bipolar plate |
| c | capillary pressure or cathode |
| cell | fuel cell |
| c,end | cathode bipolar plate |
| c,gdl | cathode gas diffusion layer |
| CL | catalyst layer |
| cond | condensation |
| d | dissolved water |
| d-g | dissolved water to vapor water |
| d-l | dissolved water to liquid water |
| e | electronic |
| eff | effective |
| equil | equilibrium |
| evap | evaporation |
| EOD | electro-osmotic drag |
| g | gas phase |
| g-l | vapor to liquid water |
| GDL | gas diffusion layer |
| $H_2$ | hydrogen |
| in | inlet |
| ion | ionic |
| l | liquid water |
| m | mass or source term |
| mem | membrane |
| $O_2$ | oxygen |
| out | outlet |
| pc | phase change heat (for source term) |
| rev | reversible |
| ref | reference state |
| rib | solid rib |
| sat | saturation |
| u | momentum (for source term) |
| v | water vapor |

## Abbreviations

| | |
|---|---|
| BD | back diffusion |
| BP | bipolar plate |
| CFRFF | composite foam-rib flow field |
| CL | catalyst layer |
| CPD | concentration permeation diffusion |
| CRFF | conventional rib flow field |
| EOD | electro-osmotic drag |
| FEM | finite element method |
| GDL | gas diffusion layer |
| GFC | gas flow channel |
| MFR | metal foam rib |
| MPL | microporous layer |
| PEM | proton exchange membrane |
| PEMFC | proton exchange membrane fuel cell |

# 1. Introduction

The proton exchange membrane fuel cell (PEMFC) is a high-efficiency energy conversion unit within the field of hydrogen energy utilization [1-6]. A fuel cell typically consists of several essential components, namely a proton exchange membrane (PEM), catalyst layers (CLs), microporous layers (MPLs), gas diffusion layers (GDLs), gas flow channels (GFCs), and bipolar plates (BPs). The GDL is a hydrophobic and porous structure positioned between the BP and CL, and it can support the CL and offer protection to both the PEM and CL [7-10]. It mainly provides a diffusion pathway for the transport of reactant gases, uniformly disperses the reactant gases so that the reactant gases diffuse to reaction sites more evenly, and provides a drainage pathway for product water [11-14]. It is electric conductive and can provide a route for electron transfer [15]. Hence, the GDL assumes a pivotal role in membrane electrode assemblies and significantly influences the overall performance of PEM fuel cells.

Given the aforementioned functions played by the GDLs, it becomes imperative to take into account various aspects of GDLs during the design and optimization of fuel cells. These aspects include parameters such as thickness, porosity, water wettability, pore size distribution, electron conductivity, thermal conductivity, and reactant transfer characteristics [16, 17]. Extensive effort has been dedicated to investigating the impact of GDL transport features on the performance of fuel cells [18-25]. For example, Xu et al. [23] utilized a 1D non-isothermal two-phase model to investigate the impact of temperature distribution on the water transport inside porous medium layers, specifically the GDL and MPL. Carcadea et al. [22], by numerical simulating a three-dimensional (3D) non-isothermal anisotropic PEM fuel cell, examined the effects of a porosity gradient within the GDL. Their results indicated that a progressive rise in porosity from the CL to the GFC resulted in more significant improvements in cell performance when compared to the traditional uniform porosity arrangement. Chang et al. [26] conducted a numerical study to evaluate the transport characteristics of the GDL. Their results revealed that enhancing the hydrophobic characteristics of the GDL led to a notable enhancement in the capability of removing liquid water. Kong et al. [27] introduced a new design of a multilayer GDL structure, containing a double-layer GDL and a single-layer MPL. They investigated the impact of the GDL's porosity and hydrophobicity on its capability to remove water. The results suggest that, compared to a single-layer GDL, the GDL's capability of removing liquid water can be enhanced by using a double-layer GDL with reduced porosity or higher hydrophobicity near the MPL. Furthermore, this design leads to a reduction in the remaining volume of liquid water within the porous medium.

To further enhance the GDL transport capability, researchers have proposed many structural optimizations based on conventional rib flow fields (CRFFs) designs for fuel cells. Yin et al. [28] introduced a new GDL design featuring elliptical grooves, which significantly enhances the velocity of reactants and products in comparison to conventional GDL. The mass fraction of water exhibited a reduction, while the mass fraction of oxygen showed a rise in the GDL when the depth was increased from 100 to 300 μm. Rectangular GDLs have been reported to experience a decline in oxygen transfer efficiency in the under-rib region of CRFF design. To tackle this concern, Son et al. [29] devised a GDL with rib-shaped structures. This GDL design was implemented in PEMFCs with various types of cathode channels. The performance of this GDL was found to be superior to that of the conventional GDL with the same cathode channel types across all voltage conditions. He et al. [30] propose an integrated GDL with a micro-tunneled rib and a wavy channel. This design aims to enhance cell performance by leveraging the transition flow effect caused by the wavy channel and the superior water removal capabilities of the micro-tunneled rib.

The thickness of GDL ($h_{GDL}$) is considered to be a critical parameter in the characterization of GDLs [31-33]. Xia et al. [34] developed a computational model that takes into account non-isothermal conditions and 3D geometry to investigate the impact of $h_{GDL}$ on the flow distribution in high-temperature PEMFCs (HT-PEMFCs). The results suggest that an augmentation in $h_{GDL}$ results in improved homogeneity of flow throughout the flow field. Additionally, the diffusion flux decreases as $h_{GDL}$ increases, and this decrease is further influenced by an increase in the GDL porosity. The hydrodynamics of liquid water in GDLs was investigated by Jeon [35] employing the multiphase lattice Boltzmann method (LBM) to simulate the impact of $h_{GDL}$. The results indicate that as $h_{GDL}$ increases, there is a rise in liquid water saturation, which leads to the obstruction of pores within the porous medium and results in an elevated resistance to oxygen transfer. Consequently, a reduction in the concentration of oxygen occurs at the reaction site within the cathode CL. Decreasing $h_{GDL}$ can offer benefits in enhancing the water removal efficiency from the porous medium. This is because a thinner GDL possesses a shorter pathway for the removal of liquid water. Nevertheless, the elimination of water films adhered to the ribs during the operation of a fuel cell is a challenge. Lee et al. [36] utilized synchronous X-ray imaging to study the impact of $h_{GDL}$ on water management with carbon paper GDLs coated with MPLs. They visualized the process of liquid water buildup within the PEM fuel cells. The results suggest that fuel cells with thinner carbon paper layers exhibit enhanced cell performance and reduced levels of liquid water saturation when subjected to high current densities. The impact of $h_{GDL}$ on the distribution of local current density was investigated by Sun et al. [37] through a 2D numerical model. The findings indicate that the influence of $h_{GDL}$ on the localized reaction rate is via the interplay between oxygen transport and electron conduction. The numerical investigation by Jeng et al. [38] shows that an increase in $h_{GDL}$ results in a reduction in cell performance, particularly when the porosity is low. Nevertheless, in cases where the porosity of the GDL is elevated and $h_{GDL}$ undergoes variations, the cell performance reaches an optimal value. The impact of $h_{GDL}$ on the performance of cells was explored by Chun et al. [33] using 1D numerical simulations. The findings indicate that as $h_{GDL}$ grows, there is a decrease in the concentration of oxygen within the CL, an increase in the saturation of liquid water, and a decline in the overall cell performance. Zhu et al. [39] investigated the effect of $h_{GDL}$ on steady-state cell performance. The results indicate the presence of an optimal $h_{GDL}$ (100 μm) of the cathode GDL that yields favorable steady-state cell performance. The findings of existing research on CRFF fuel cells indicate that the performance of cells firstly improves and subsequently declines with increasing $h_{GDL}$. These results imply that there is an optimum $h_{GDL}$ that results in the best performance of the fuel cell. Huo et al. [40] conducted a comparative analysis of cell performance utilizing metal foam flow field (MFFF) structures and those employing CRFF structures. The porous MFFF structure can improve the transport of water and gas at the interface between the GDL and metal foam. This expedites the dehydration process of the cathode in PEMFCs. Additionally, the implementation of porous electrodes facilitates more homogeneous electron transport, thereby enabling the utilization of thinner GDLs in MFFF fuel cells.

The utilization of the composite foam-rib flow field (CFRFF) design can modify the gas flow dynamics and improve the under-rib oxygen transfer and water removal processes. Consequently, this can lead to enhanced cell performance without an increase in the pumping power [41]. While the effect of $h_{GDL}$ on CRFF design has been extensively investigated, there is a notable absence of research on the effect of $h_{GDL}$ on fuel cells with the CFRFF design, and the detailed mechanisms governing multiphase transfer in the flow field are still unclear. Hence, this study investigates the effect of $h_{GDL}$ on fuel cell performance with the CFRFF design through numerical modeling. In this study, a 3D numerical model is constructed to examine the effects of $h_{GDL}$ on the performance of fuel cells with

CRFF and CFRFF designs. This study analyzes the effect of changing $h_{GDL}$ and rib width on the cell performance based on the proposed CFRFF structures and finds a new trend different from that of the CRFF structures, which can provide a reference for the further optimization of PEMFC cells.

## 2. Fuel cell geometries and numerical method

### 2.1 Fuel cell geometries

Figure 1a depicts a 3D geometric model of a fuel cell with a CFRFF structure design in the cathode flow field in this study. The solid rib is partially replaced with nickel metal foam, leading to a metal foam rib (MFR), and the bottom of the MFR remains supported by solid ribs. The height of the MFR is set to the half of the solid rib ($H_{MFR} = 0.5H_{rib}$), while the total height remains unchanged ($H_{rib} = H_{total} = 0.5H_{rib} + H_{MFR}$), including the height of the solid ribs that have not been replaced with metal foam. For comparative analysis, a geometric model is created in which the cathode is comprised of CRFF, as illustrated in Figure 1b. Table S1 in the Supplementary Material provides the parameters employed in the simulation model [20, 42-46]. To study the influence of $h_{GDL}$, a systematic variation of the GDL thickness on the cathode and anode sides, $h_{c,GDL}$ and $h_{a,GDL}$, is conducted, respectively, in the range of 60 to 290 μm. When $h_{GDL}$ on the cathode (anode) is varied, $h_{GDL}$ on the anode (cathode) remains constant at 210 μm, if not explicitly stated otherwise. The model's coordinate origin is situated at the midpoint of the cathode plate within the cross-sectional area of the inlet, as illustrated in Figure 1.

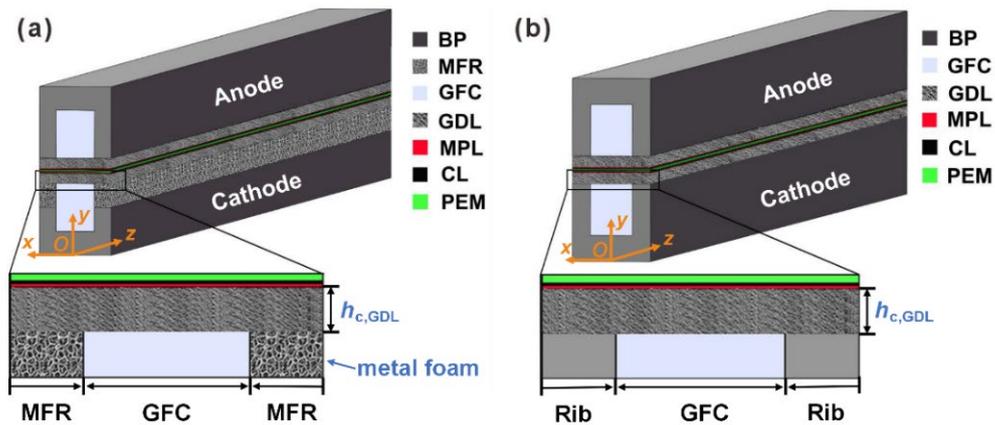

Figure 1 3D geometry of (a) CFRFF fuel cell and (b) CRFF fuel cell.

### 2.2 Model simplifications and assumptions

The present study employs a numerical model which simplifies the PEMFC by making a series of assumptions. The PEMFC is analyzed in the steady-state case, with the assumption that the gravitational influences on transport phenomena are negligible. The assumption of isotropic homogeneity is made for the porous medium (MFRs, GDLs, MPLs, and CLs). The gas flow in the cathode and anode flow fields are laminar and follow the Fick's diffusion law and the ideal gas law. The gases cannot pass through the PEM because of their impermeability. A single-phase mist flow exists within the GFC because of the quick elimination of liquid water from the GFC-GDL interface.

**2.3 Governing equations and boundary conditions**

The numerical model is 3D, multiphase, and non-isothermal. This section provides an overview of the physical processes, including electrochemical reactions, mass transfer, two-phase flow, electron/ion conduction, phase change, and heat conduction. The detailed forms of the governing equations and the coupling relationship among the equations are shown in Figure 2.

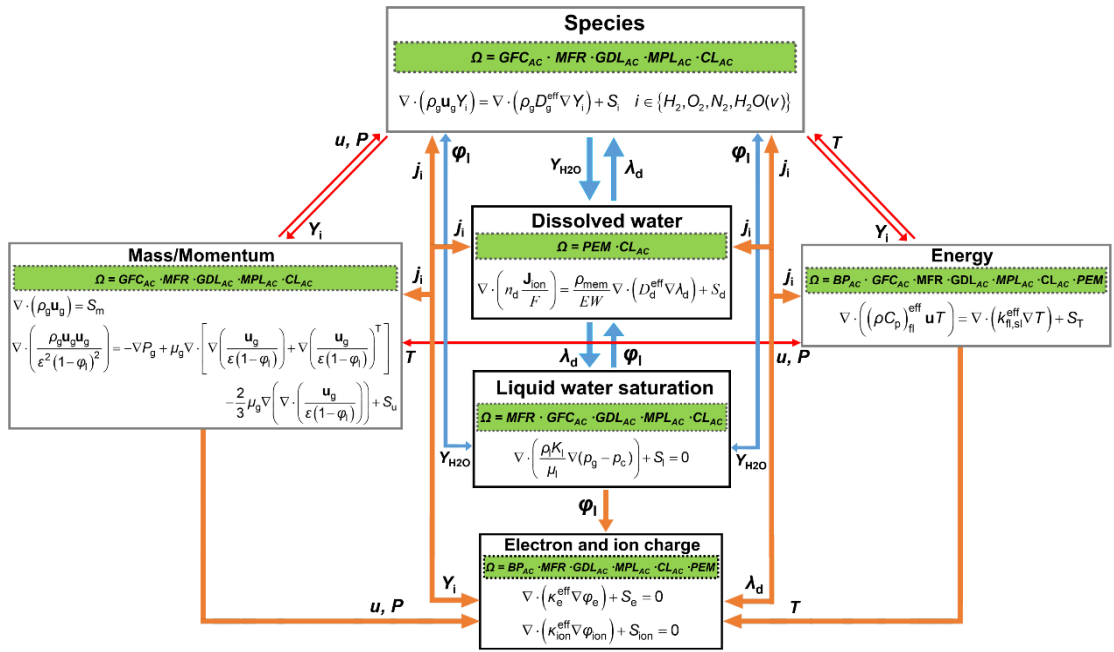

Figure 2 Conservation equations and coupling relationship among the computational domains. The red line indicates the transport relationship between the reactants; the blue line represents the transformation between liquid water, dissolved water, and vapor; and the orange line displays the coupling relationship between the electric current and other physical fields.

Figure 3 displays the boundary condition settings utilized in the 3D model of the PEMFCs. Table S2 in Supplementary Material provides a comprehensive summary of the relationships among phase change, gas diffusion, charge transfer, dissolved water transfer, and electrochemical reactions [46-54]. The source terms for the conservation equations are provided in Table S3 in Supplementary Material [11, 55-59].

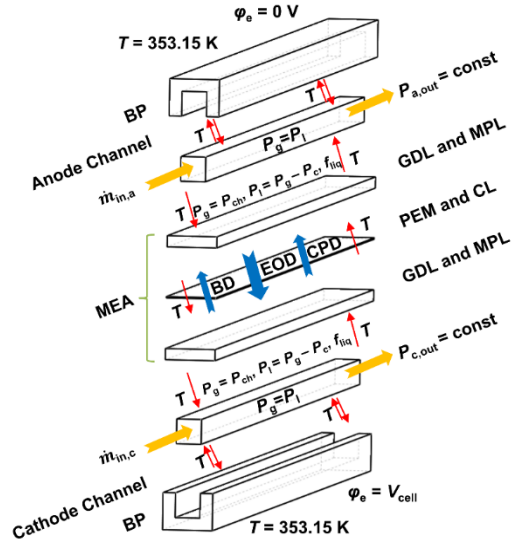

Figure 3 Boundary conditions for the governing equations in the simulation.

**2.5 Validation of the simulation**

The 3D conservation equations of PEMFCs were solved in COMSOL Multiphysics by employing the finite element method (FEM). To determine the accuracy of the simulation, we examine the current densities at 0.6 V and 0.4 V that are produced by running the model with varying the number of grid elements, namely 4160, 28050, 74800, 110000, and 158400, as shown in Figure 4a. The mesh with 110,000 cells demonstrates grid independence, thereby making it appropriate for subsequent simulations. A comparison is conducted between the simulation results and the corresponding experimental data, which was obtained at different ionomer volume fractions ($\omega = 0.22$ and 0.27). The experimental conditions described in Ref. [60] were incorporated into the PEMFC model's setting parameters and operating conditions. As depicted in Figure 4b, the numerical results for the 3D model show that they agree well with the results of the experiments.

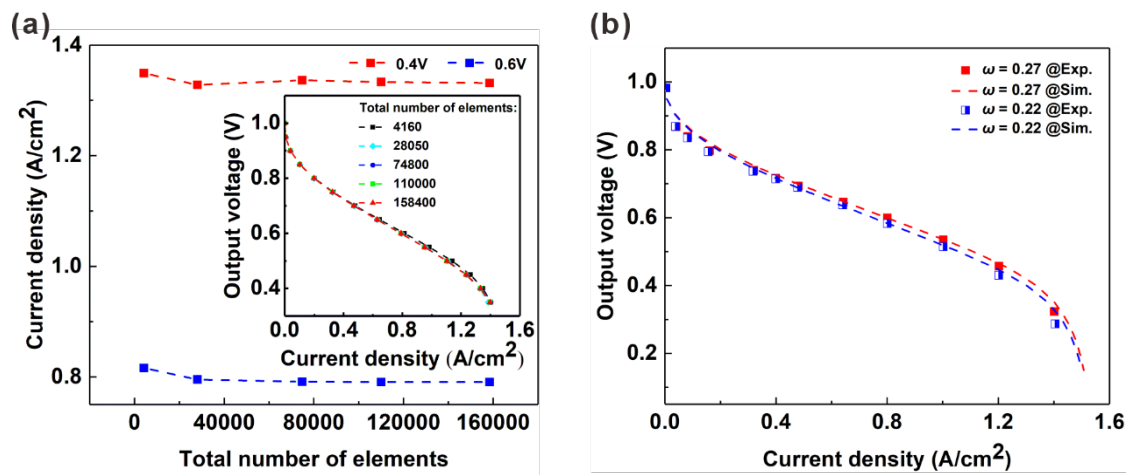

Figure 4 (a) Grid independence study at various grid densities. (b) Comparison of experimental and simulation results for different ionomer volume fractions.

## 3. Results and discussion

### 3.1 Comparison of CFRFF and CRFF fuel cells under different cathode GDL thicknesses

We firstly explore the influence of the cathode GDL thickness $h_{c,GDL}$ on the performance of the cell with the CRFF and CFRFF designs, as depicted in Figures 5a and b, respectively. For the CRFF design, as $h_{c,GDL}$ increases, the cell performance initially increases, followed by a decrease, suggesting the presence of an optimal value of $h_{c,GDL}$. The CRFF design exhibits the maximum peak power density when $h_{c,GDL}$ = 130 μm, as depicted in Figure 5a. In contrast, for the CFRFF design, the performance of the cell demonstrates an increasing trend as $h_{c,GDL}$ decreases, as depicted in Figure 5b. Furthermore, under identical operating conditions, the performance of the cell with the CFRFF design surpasses that of the CRFF design throughout the whole parameter range. For the CRFF and CFRFF designs with different $h_{c,GDL}$, the corresponding polarization curves almost overlap at high output voltages (0.8–1.0 V) and medium output voltages (0.5–0.8 V); however, the polarization curves for both CRFF and CFRFF design at low output voltages (0.1–0.4 V) with different $h_{c,GDL}$ show obvious differences. At low operating voltages, the limiting current density of the CFRFF fuel cell shows an increasing trend with decreasing $h_{c,GDL}$. Hence, it can be judged that the influence on ohmic polarization loss and activation polarization loss is small, and the influence of different $h_{c,GDL}$ on cell performance is mainly via concentration polarization. Hence, we next conduct an in-depth analysis of the internal mechanism from the perspectives of mass transfer and water removal. The GDL plays the role of a crucial conduit for mass transfer and water elimination. The primary objective of this section is to investigate the influence of $h_{c,GDL}$ on the performance of fuel cells, with a specific focus on two critical factors: oxygen transfer and water removal.

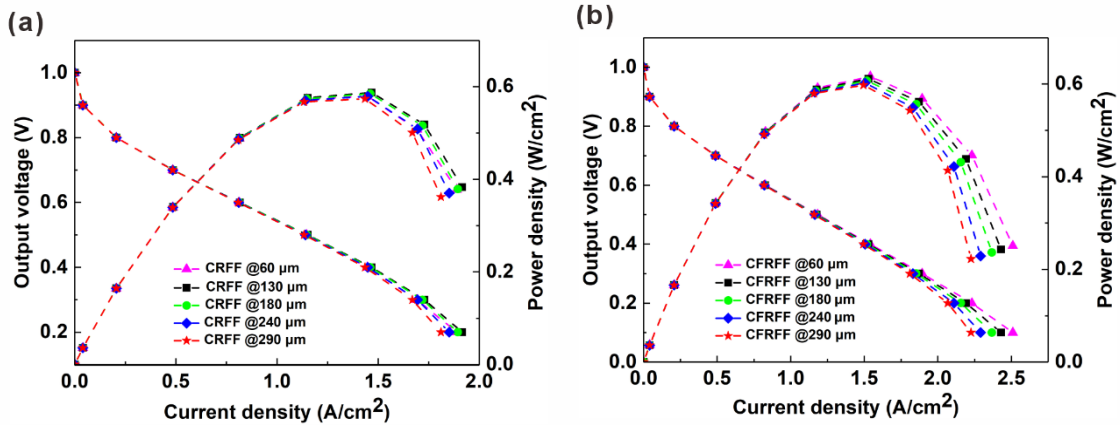

Figure 5 Impacts of $h_{c,GDL}$ on the cell performance for (a) CRFF and (b) CFRFF fuel cells.

First, we analyze the factors contributing to the variance in cell performance at different $h_{c,GDL}$ for the two flow field designs. This analysis focuses on the perspective of mass transfer, as depicted in

Figure 6. A strong correlation exists between the performance of the fuel cell and the oxygen concentration that is transferred to the site of the three-phase electrochemical reaction in the CL via the flow field. At 0.6 V, as $h_{c,GDL}$ decreases, the average concentration of oxygen for the fuel cell with CRFF design increases first and then decreases, while the average concentration of oxygen for the fuel cell with CFRFF design increases linearly. At 0.2 V, as $h_{c,GDL}$ decreases, the average oxygen concentration in both CRFF and CFRFF fuel cells increases, as shown in Figure 6a and b.

Additional analysis is conducted to investigate the cause of the fluctuation in the average concentration of oxygen at 0.2 V and 0.6 V, with a focus on the distribution of oxygen concentration. The reactant gas transfer efficiency of the CRFF design is influenced by the solid rib structure. The accumulation of under-rib liquid water has an impact on the under-rib oxygen concentration of fuel cells with CRFF designs. At 0.6 V, the under-rib oxygen concentration consistently drops when $h_{c,GDL}$ decreases, while it consistently increases under the channel. Notably, at the MPL-CL interface with $h_{c,GDL} = 130$ μm, the average oxygen concentration reaches its highest value, as depicted in Figure 6c. At 0.2 V, with the decrease in $h_{c,GDL}$, the increase in electrochemical reaction intensity produces much liquid water under the ribs, which seriously blocks the oxygen transfer. Meanwhile, as shown in Figure 6d, the gas transport path under the channel will be shortened, which increases the concentration of oxygen under the channel. As a result, the average concentration of oxygen in the CRFF design increases with decreasing $h_{c,GDL}$ at 0.2 V. In contrast, the CFRFF design exhibits an increase in both the under-channel and under-rib oxygen concentrations as $h_{c,GDL}$ decreases, as shown in Figure 6e and f.

As depicted in Figure 7a, the primary factor contributing to the reduction in under-rib oxygen concentration in the CRFF design is the buildup of under-rib liquid water because of the existence of solid ribs. This phenomenon is particularly pronounced when $h_{c,GDL}$ decreases. In the CRFF design, liquid water is easily accumulated in the under-rib region. With the decrease in $h_{c,GDL}$, although the water removal path is shortened, it increases the buildup of under-rib liquid water, further increasing the barrier to under-rib oxygen transport. It can also be seen in Figure 7a that severe flooding occurs in the under-rib region at $h_{c,GDL} = 60$ μm. The excessive liquid water in the reaction region results in a small reaction area and a low catalyst utilization efficiency, making the utilization of reaction gases less efficient and thus limiting the performance of the fuel cell with the CRFF design. However, the distribution of liquid water saturation in the CFRFF design exhibits higher uniformity, and the decrease in $h_{c,GDL}$ further enhances the capability of removing liquid water in the GDL, showing a decreasing trend of liquid water saturation as $h_{c,GDL}$ decreases, as depicted in Figure 7b.

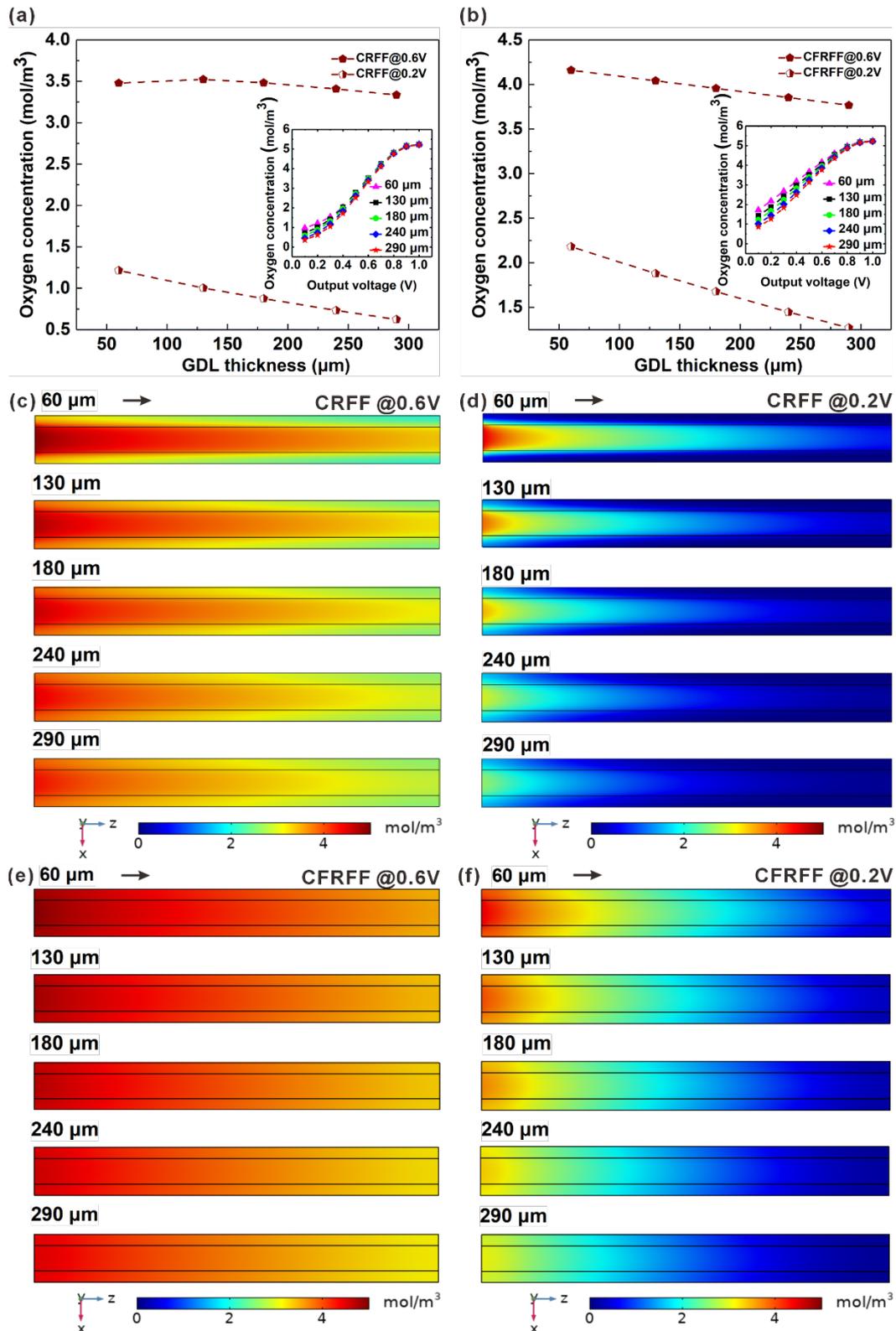

Figure 6 (a,b) Effect of $h_{c,GDL}$ of (a) CRFF and (b) CFRFF fuel cells on average oxygen concentration at different operating voltages at the MPL-CL interface. (c-f) Oxygen concentration distribution in the CRFF design at (c) 0.6 V and (d) 0.2 V and in the CFRFF design at (e) 0.6 V and (f) 0.2 V under different $h_{c,GDL}$ at the MPL-CL interface.

Furthermore, Figures 7c and 7d illustrate the average saturation of liquid water under various operating voltages. The saturation of liquid water in the CFRFF design is found to be lower for different $h_{c,GDL}$ values at varying operating voltages compared with the CRFF design. This can be attributed to the enhanced capability of removing water in the CFRFF design. The cell performance of PEM fuel cells with the CRFF and CFRFF designs is determined by the interplay between oxygen transport and water drainage.

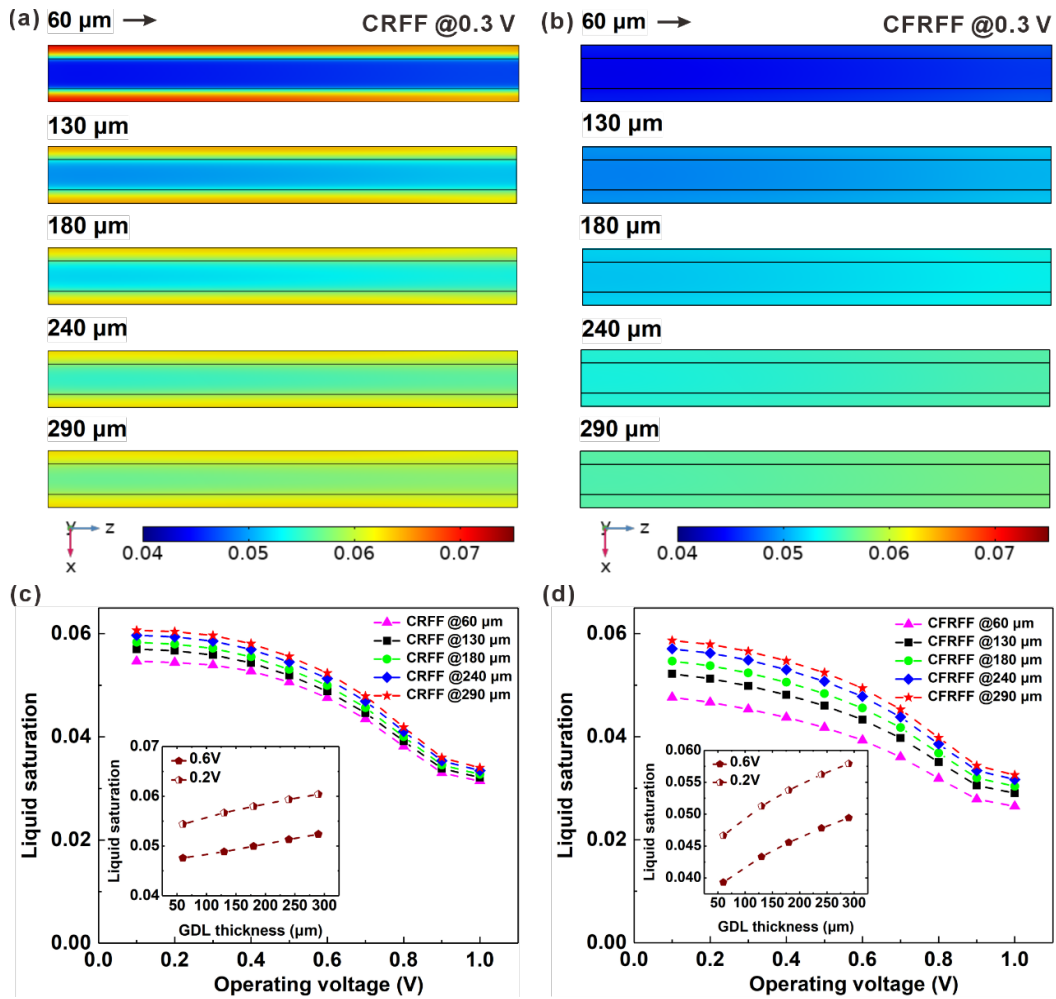

Figure 7 (a,b) Saturation of liquid water distribution of (a) CRFF and (b) CFRFF designs under different $h_{c,GDL}$. (c,d) Effect of $h_{c,GDL}$ of fuel cells of (c) CRFF and (d) CFRFF designs on the average saturation of liquid water under different operating voltages.

Following the above analysis of the distribution of oxygen concentration and liquid water saturation in the in-plane direction, we next analyze the distribution in the through-plane direction, as shown in Figure 8. As the electrochemical reaction proceeds, the oxygen concentration is consistently depleted. The CRFF and CFRFF fuel cells both exhibit a decline in oxygen concentration along the direction of flow but have differences for different GDL thicknesses. The oxygen concentration of the

CFRFF fuel cell under the rib is greater than that of the CRFF fuel cell in the through-plane direction. Additionally, the liquid water saturation of the CFRFF fuel cell under the rib is lower than that of the CRFF fuel cell at various GDL thicknesses. These differences can be attributed to the variation in the paths of oxygen transport and water removal. For the CRFF fuel cell, significant under-rib flooding can be observed at 60 μm, as shown in Figure 8c.

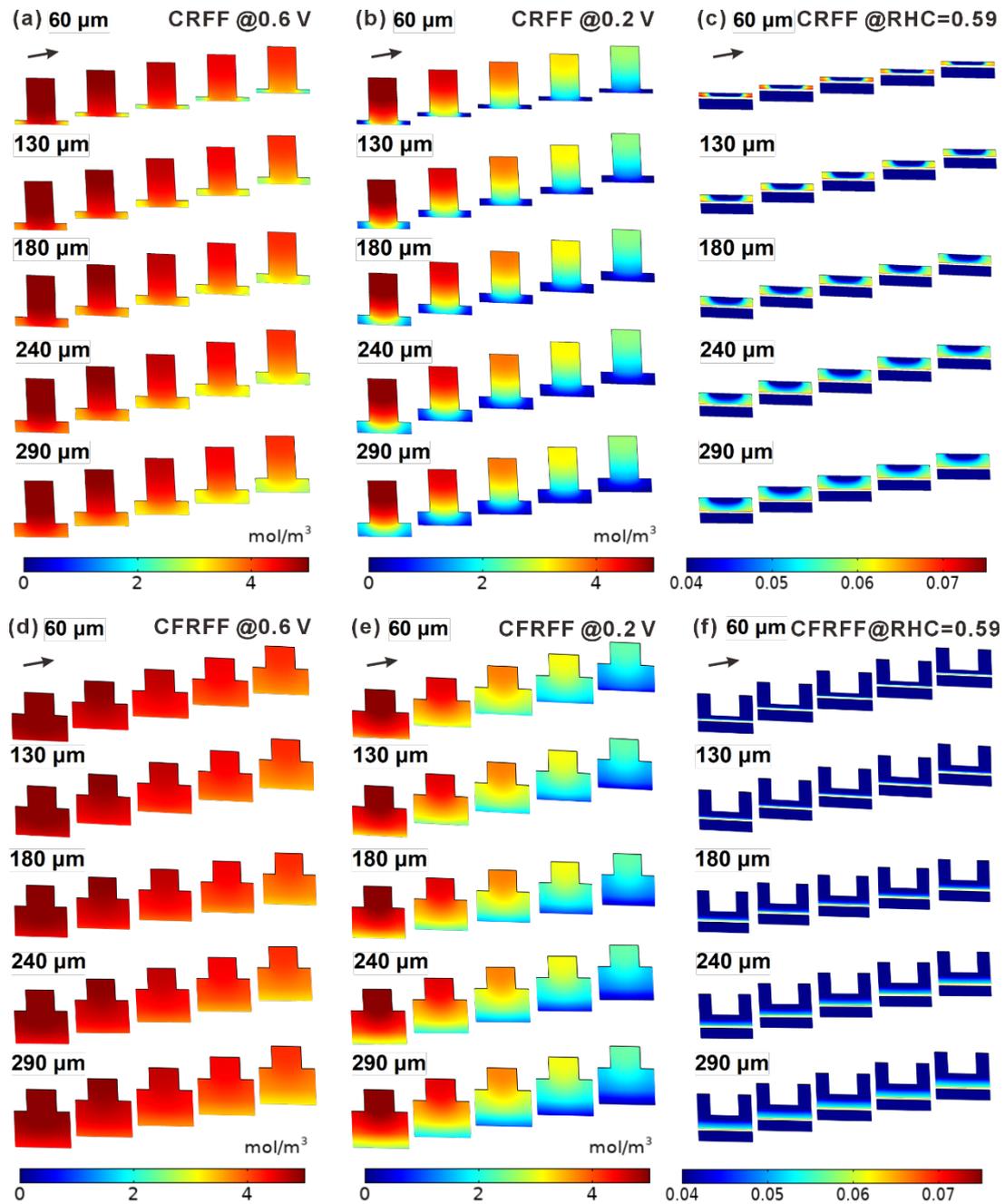

Figure 8 (a-c) Oxygen concentration distribution at (a) 0.6 V and (b) 0.2 V and (c) saturation of liquid water distribution under RHC = 0.59 (inlet relative humidity in the cathode) of the CRFF fuel cell under different $h_{c,GDL}$. (d-f) Oxygen concentration distribution at (d) 0.6 V and (e) 0.2 V and (f) saturation of liquid water distribution under RHC = 0.59 of the CFRFF fuel cell under different $h_{c,GDL}$.

The current density's magnitude and distribution are influenced by multiple parameters, such as the oxygen concentration, electron and ion potentials, liquid water saturation, and dissolved water content and distribution. Therefore, these parameters can together function as indicators of cell performance. The trends of the magnitude and distribution of the current density against the varying $h_{c,GDL}$ are depicted in Figure 9. For the CRFF design, the variation in the distribution of local current density aligns with the variation in oxygen concentration at a potential of 0.6 V when $h_{c,GDL}$ is reduced. This indicates that the effect on the density of current is dominated by the concentration of oxygen. Therefore, at $h_{c,GDL}$ = 130 μm, the cell current density is higher than others overall, as shown in Figure 9a. At 0.2 V, the under-channel current density increases with the decrease of $h_{c,GDL}$, which is mainly affected by the under-channel oxygen concentration. The under-rib current density reduces as $h_{c,GDL}$ decreases, mainly because of the flooding effect under the rib. At $h_{c,GDL}$ = 60 μm, the under-rib current density is seriously affected by the under-rib flooding, which leads to a much smaller current density under the ribs than that under the channels. An excessively thin GDL will result in an inhomogeneous reaction rate distribution, a low reaction efficiency, and a low reaction gas utilization rate relative to $h_{c,GDL}$ = 130 μm. Therefore, the cell performance at $h_{c,GDL}$ = 130 μm is also better than at the other GDL thicknesses for the CRFF design at 0.2 V, as shown in Figure 9b.

For the CFRFF design, the density of local current at the cathode MPL-CL interface exhibits an increase in magnitude as $h_{c,GDL}$ declines, regardless of whether the operating voltage is high or low, as depicted in Figure 9c and d. This is consistent with the trend in Figure 5b, which illustrates the behavior of the polarization curves. This is the result of reduced concentration polarization due to the shorter paths of oxygen transport and liquid water removal.

In summary, in the CRFF design, because of the existence of the solid rib, $h_{c,GDL}$ has an important influence on the under-rib oxygen transfer capability and the cell performance. A thinner cathode GDL can improve the gas supply capability, while the under-rib liquid water is easier to accumulate, resulting in higher mass transfer resistance and lower reaction efficiency. In a thicker cathode GDL, the reactant transfer performance is impaired due to the longer transport path in the GDL. Therefore, there is an optimal value of $h_{c,GDL}$ for the CRFF design, which is $h_{c,GDL}$ = 130 μm in this study. The concentration of oxygen in the CL in the CFRFF design is higher, and the distribution of reactants is more uniform than that of the CRFF design. Hence, the performance of the CFRFF design is better under different $h_{c,GDL}$. Additionally, the performance further improves as $h_{c,GDL}$ of the CFRFF design decreases, which is distinct from the CRFF design.

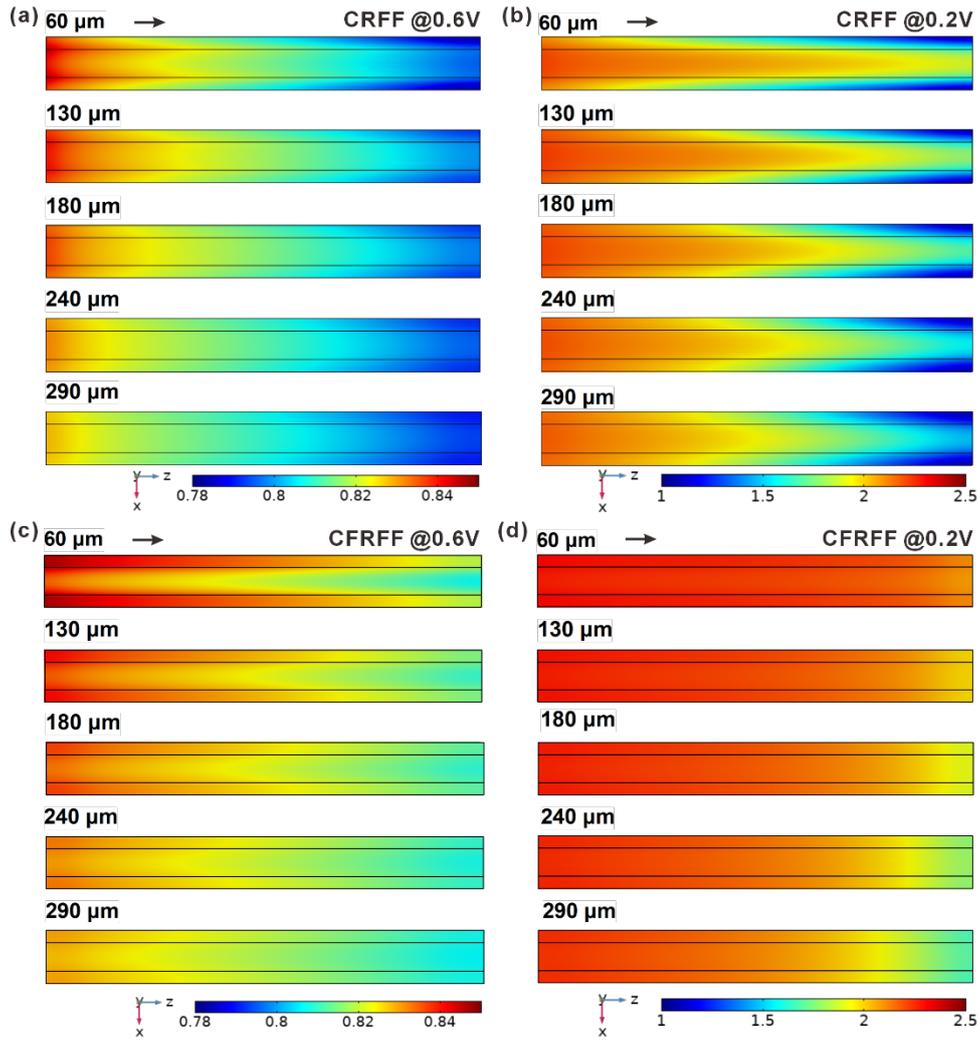

Figure 9 Distribution of current density of the CRFF fuel cell at (a) 0.6 V and (b) 0.2 V and the CFRFF fuel cell at (c) 0.6 V and (d) 0.2 V under different $h_{c,GDL}$.

**3.2 Impact of cathode rib width for CFRFF and CRFF fuel cells with thin cathode GDLs**

The cathode rib width is also a significant variable affecting the capabilities of under-rib oxygen transfer and under-rib water removal, especially at thin GDLs. Therefore, the cell performance of the CRFF and CFRFF designs, with a thin GDL (i.e., $h_{c,GDL}$ = 60 μm, where the cell performance is seriously affected by the solid rib structure at this condition), is studied under different rib-width ratios ($\phi_{rib}$) in the cathode, as shown in Figure 10a and b. Here, the rib-width ratio is defined as the ratio of the width of the solid rib $W_{rib}$ to the total width of the solid rib and the gas flow channel ($W_{rib} + W_{channel}$), i.e.,

$$\phi_{rib} = \frac{W_{rib}}{W_{rib} + W_{channel}}$$

In this study, the total width of the solid rib and the gas flow channel is maintained constant ($W_{rib} + W_{channel}$ =1.5 mm). Compared with oxygen at the cathode, considering that hydrogen is transported much faster at the anode and the variation in the anode GDL has a weak effect on the performance of the fuel cell, the anode GDL thickness is kept at $h_{a,GDL} = 210$ μm. As depicted in Figure 10a, the density of the peak power and the density of the limiting current in the CRFF design are reduced with increasing $\phi_{rib}$, and the cell performance decreases. However, the density of peak power for the CFRFF design increases with increasing $\phi_{rib}$, as shown in Figure 10b. Next, this change is analyzed in terms of the change in under-rib oxygen transfer and under-rib water removal capacity.

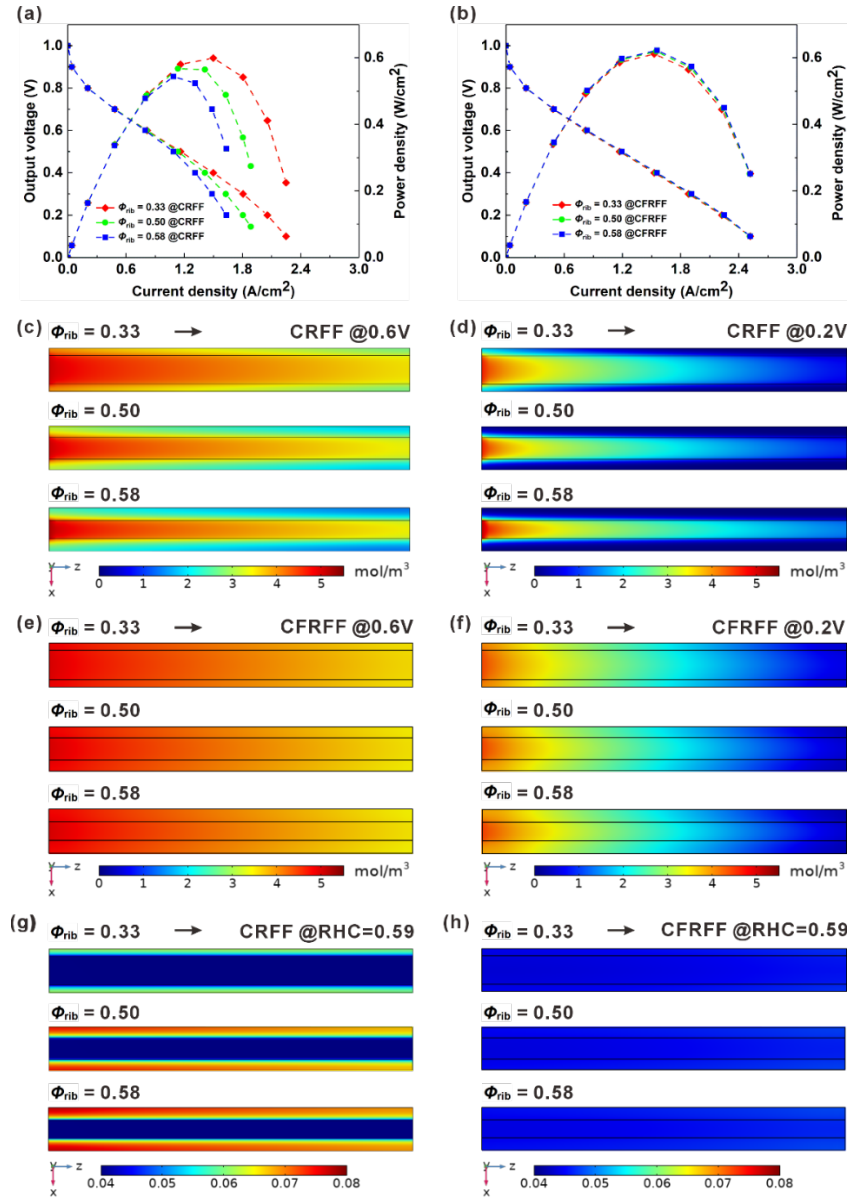

Figure 10 (a, b) Comparison of cell performance between fuel cells of (a) CRFF and (b) CFRFF designs under different $\phi_{rib}$. (c-f) Influence of $\phi_{rib}$ on the distribution of oxygen concentration in the CRFF design at (c) 0.6 V and (d) 0.2 V and in the CFRFF design at (e) 0.6 V and (f) 0.2 V. (g, h) Impact of $\phi_{rib}$ on the distribution of liquid water saturation in (g) CRFF and (h) CFRFF designs.

For the CRFF design, at a thin cathode GDL, the presence of the solid rib structure produces a severe accumulation of under-rib liquid water, which increases the resistance to under-rib reactant gas transfer and reduces the performance of the fuel cell, as depicted in Figure 10c and d. The increase in $\phi_{rib}$ enlarges the under-rib reactant gas transfer region, further increasing the difficulty of under-rib oxygen transfer, which is the primary factor contributing to the reduced performance of the fuel cell. However, for the CFRFF design, due to the capability of the MFR structure to change the gas transport pattern, which avoids the effect of solid ribs in the CRFF design on mass transfer, the under-MFR oxygen concentration does not change appreciably with $\phi_{rib}$ increases, as shown in Figure 10e and f. In addition, with the increase in $\phi_{rib}$ in the CRFF design, the under-rib liquid water also increases significantly, and the water accumulation region under the ribs increases compared with the CFRFF design, as shown in Figure 10g and h.

The oxygen concentration distributions of the CRFF and CFRFF designs in the under-rib and under-channel regions are further compared, as shown in Figure 11a-b. At 0.6 V, for the CRFF design, as the rib width increases on both sides of the channel, the concentration of oxygen decreases both in the under-channel and under-rib regions. In particular, the concentration of oxygen reduces rapidly in the under-rib region. At 0.2 V, the oxygen concentration under the channel no longer decreases significantly with increasing $\phi_{rib}$, while the concentration of oxygen in the under-rib region still decreases sharply with increasing rib width. The drop in oxygen content is mostly caused by an increase in the saturation of liquid water in the under-rib porous medium. As shown in Figure 11c, for the CRFF design, at 0.2 V and 0.6 V, the under-rib liquid water increases dramatically with increasing $\phi_{rib}$ and accumulates under the ribs, thus clogging the porous medium and increasing the concentration polarization loss. However, the concentration of oxygen and the saturation of liquid water under the rib and under the channel changed little for the CFRFF design, with increasing $\phi_{rib}$, at 0.2 V and 0.6 V, as depicted in Figure 11b and d.

The primary factor contributing to the performance variability of fuel cells in the CFRFF design is the alteration in Ohmic polarization loss due to the increase in $\phi_{rib}$, while the increase in $\phi_{rib}$ has a smaller impact on the concentration polarization loss. The dissolved water content in the ionomers increases with increasing $\phi_{rib}$ under different operation voltages, as depicted in Figure 11e, and the variance in the content of dissolved water is mainly at the cathode, as shown in Figure 11f. This can minimize the proton transfer resistance at the cathode, which decreases the Ohmic polarization loss, resulting in a 1.9% increase in the density of peak power for the CFRFF design with $\phi_{rib} = 0.58$, compared with the CFRFF design with $\phi_{rib} = 0.33$.

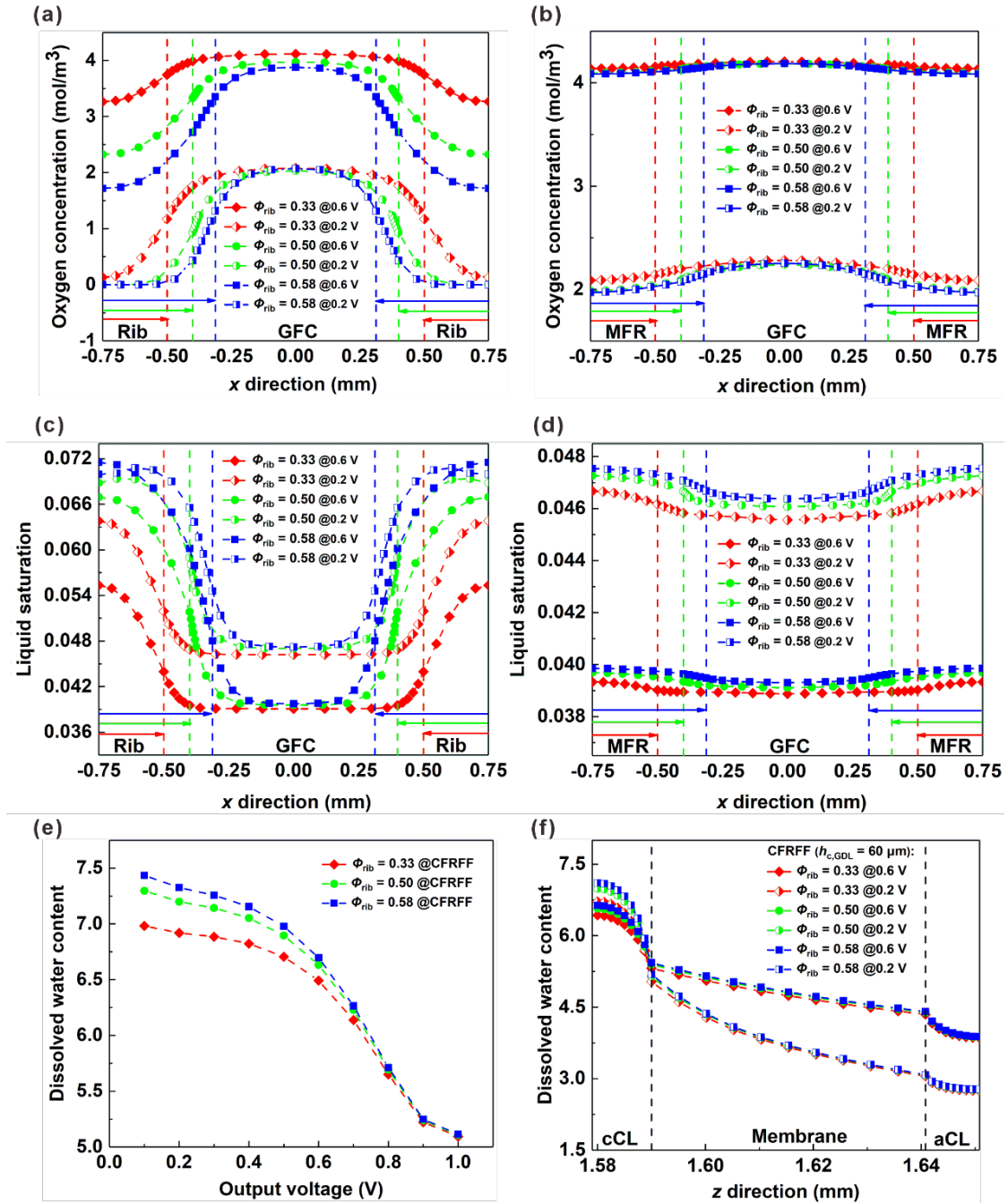

Figure 11 (a-d) Oxygen concentration distribution of (a) CRFF and (b) CFRFF designs and the liquid saturation distribution of (c) CRFF and (d) CFRFF designs along the spanwise direction (i.e., $x$ direction) at the MPL-CL interface (i.e., $y = 1.58$ mm) at the mid-point of the channel (i.e., $z = 25$ mm). The vertical dashed lines indicate the position of the interfaces between the GFC and the solid rib. (e) Average dissolved water content of CFRFF design with different $\phi_{rib}$ under different operation voltages. (f) Distribution of dissolved water content along the $y$ direction of the CFRFF design (i.e., $x = 0$ mm and $z = 25$ mm) under different $\phi_{rib}$ at 0.6 V and 0.2 V. The vertical dashed lines indicate the positions of the membrane-CL interfaces.

In conclusion, at a thin GDL thickness for the CRFF design, cell performance decreases with increasing $\phi_{rib}$. This is mostly due to the increased under-rib region's resistance to under-rib oxygen transfer, which increases concentration polarization loss. However, for the CFRFF design, with the increase in $\phi_{rib}$, its effect on the under-rib region's resistance to oxygen transport is small, but it can weaken the water removal capability, thus increasing the content of dissolved water in the cathode ionomer, which decreases the Ohmic polarization loss and thus improves the cell performance. These results suggest that at thin GDL thicknesses, $\phi_{rib}$ should be minimized for the CRFF design and slightly increased for the CFRFF design.

### 3.3 Effect of anode GDL thickness for CFRFF and CRFF fuel cells

As depicted in Figure 12, the influence of the anode GDL thicknesses ($h_{a,GDL}$) of the CFRFF and CRFF designs on the performance of fuel cells is also researched. When only $h_{a,GDL}$ is changed, $h_{c,GDL}$ remains unchanged at 210 μm. As $h_{a,GDL}$ decreases, the cell performance of the CRFF and CFRFF designs improves, but the amplitude of the improvement is small, as depicted in Figure 12a and b. When $h_{a,GDL}$ is reduced from 240 to 60 μm, the density of peak power is increased by 1.7% for the CRFF design and by 2.3% for the CFRFF design. Meanwhile, the density of limiting current changes little for both the CRFF and CFRFF designs.

The distribution of oxygen concentration in the CRFF and CFRFF designs at 0.2 and 0.6 V is shown in Figure 12c-f. For the CRFF design, under different $h_{a,GDL}$ at 0.2 or 0.6 V, the concentration of oxygen at the cathode CL-MPL interface shows little change. The CFRFF design has a similar changing trend to the CRFF design. This is because changing $h_{a,GDL}$ has a direct effect on the anode mass transfer but does not affect the cathode mass transfer capability. However, the concentration polarization loss is mainly on the cathode CL rather than on the anode side. Therefore, as $h_{a,GDL}$ reduces, the density of limiting current changes only slightly, and the effect on cell performance is mainly reflected in the Ohmic polarization loss.

As the $h_{a,GDL}$ decreases, the content of dissolved water in the ionomer increases, which can reduce the Ohmic loss, especially in the cathode CL, as shown in Figure 12g and h. This is because, as $h_{a,GDL}$ decreases, the hydrogen transfer path shortens, the hydrogen transport capability improves, and the proton transport capability improves, promoting the electro-osmotic transport of dissolved water from the anode to the cathode and further increasing the content of dissolved water in the cathode ionomer. Therefore, the peak power density of the CRFF and CFRFF designs increases with decreasing $h_{a,GDL}$.

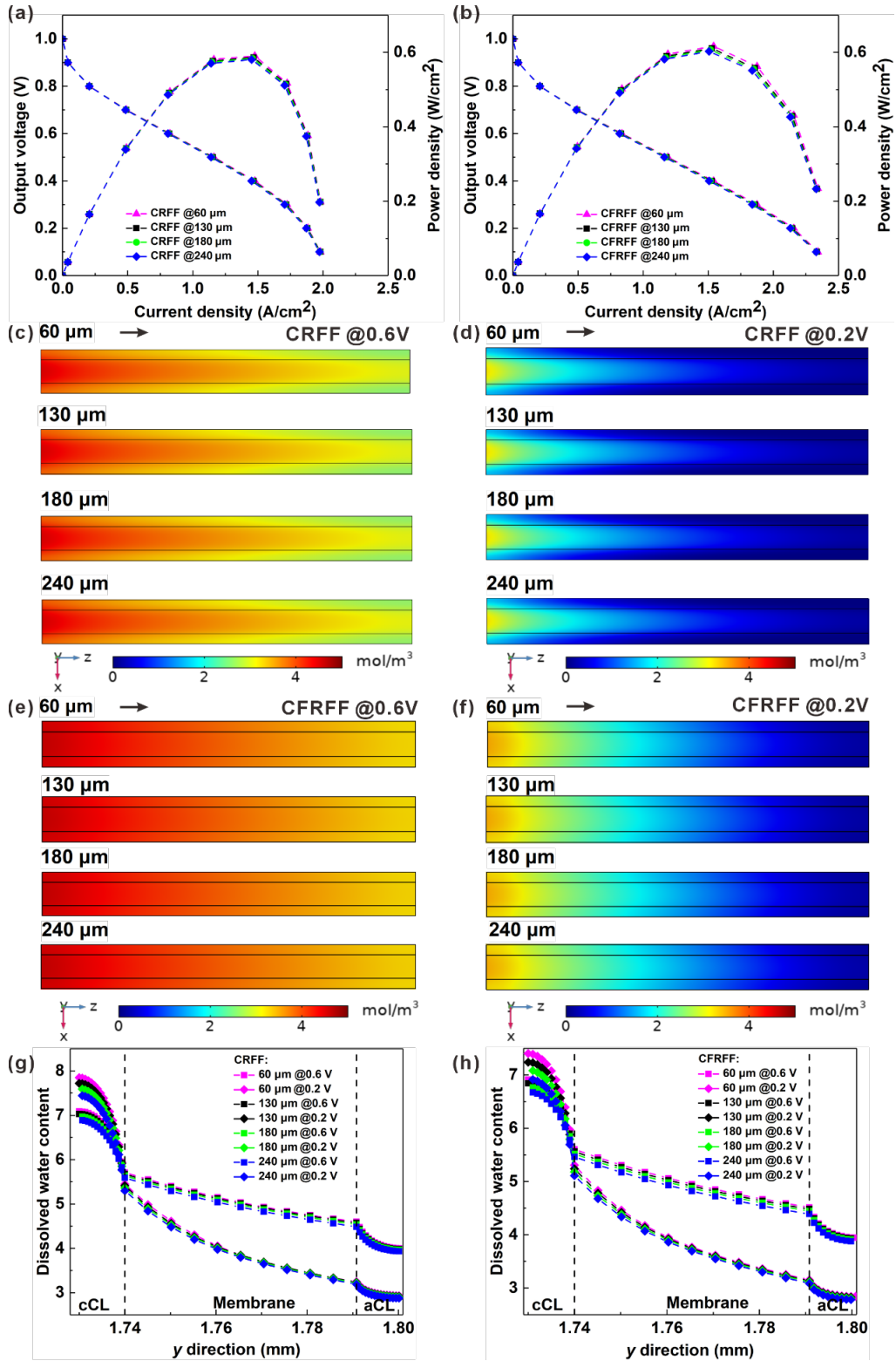

Figure 12 (a,b) Effects of $h_{a,GDL}$ on the cell performance for (a) CRFF and (b) CFRFF designs. (c-f) Effect of $h_{a,GDL}$ on the distribution of oxygen concentration in the CRFF design at (c) 0.6 V and (d) 0.2 V and in the CFRFF design at (e) 0.6 V and (f) 0.2 V. (g,h) Distribution of dissolved water content along the $y$ direction ($x$ = 0 mm, $z$ = 25 mm) of (g) CRFF and (h) CFRFF designs under different $h_{a,GDL}$ at 0.2 V and 0.6 V. The vertical dashed lines indicate the positions of the membrane-CL interfaces.

To sum up, the decrease in the anode GDL thickness $h_{a,GDL}$ enhances the hydrogen transfer capability, which further enhances the conduction of protons and increases the content of dissolved water in the ionomer, thus reducing the Ohmic polarization loss. Since there is no direct impact on the capability of oxygen transport in the cathode, the concentration polarization loss is almost unaffected.

**3.4 Effect of cathode GDL thickness for CFRFF design under different RHC**

Researchers have thoroughly investigated the impact of operating conditions on cell performance. Here, the performance of the CFRFF and CRFF fuel cells at different relative humidity in the cathode (RHC) is shown in Figure 13. At higher operating voltages, increasing the relative humidity improves the wettability of the proton exchange membrane, reducing the resistance to ion transport and increasing the peak power density. Under lower operating voltages, a rise in relative humidity leads to an elevation in the amount of liquid water saturation in the flow field. This, in turn, raises the resistance to oxygen transfer and lowers the limiting current density. The CFRFF design exhibits superior cell performance compared to the CRFF design in terms of peak power density and limiting current density, regardless of high or low relative humidity circumstances.

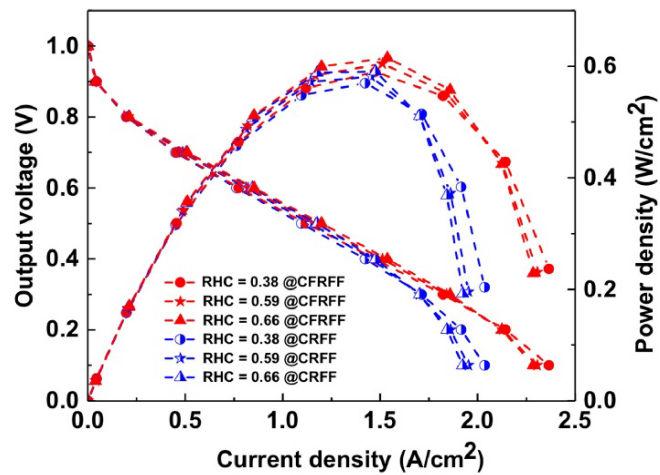

Figure 13 Effect of RHC for the performance of the CFRFF and the CRFF fuel cells ($h_{a,GDL} = h_{c,GDL} = 210$ μm).

The effect of different RHC on the performance of the CFRFF fuel cell with different $h_{c,GDL}$ is shown in Figure 14. Under different RHC, the variation in $h_{c,GDL}$ of the CFRFF fuel cell affects both the ohmic polarization and concentration polarization. Within a certain range of RHC variations, the peak power density and limiting current density decrease with the increase in $h_{c,GDL}$ under the same RHC. However, at the same $h_{c,GDL}$, the peak power density increases with the increase in RHC; and the limiting current density decreases with the increase in RHC. Hence, for the CFRFF fuel cell, it is advisable to use thinner $h_{c,GDL}$ and also to operate at higher RHC conditions.

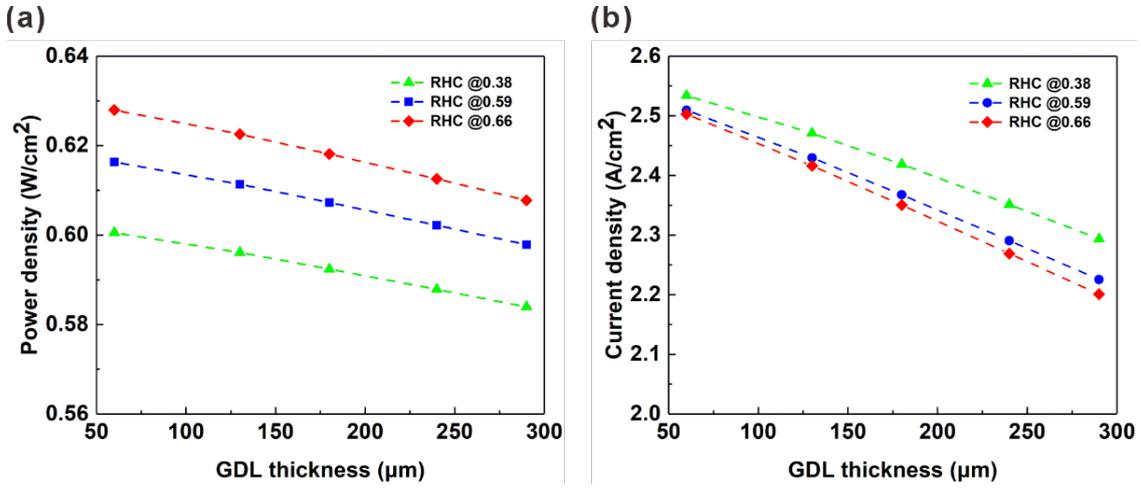

Figure 14 (a) Peak power density and (b) limiting current density for CFRFF design with different $h_{c,GDL}$ under various RHC.

## 4. Conclusions

The effect on the GDL thickness ($h_{GDL}$) in PEM fuel cells with conventional rib flow field (CRFF) and composite foam-rib flow field (CFRFF) designs is studied employing a 3D multiphase non-isothermal simulation model. The findings indicate that the CFRFF design has an improvement in cell performance as $h_{c,GDL}$ decreases, while the CRFF design has optimal performance at a moderate GDL thickness. In addition, the effects of $h_{a,GDL}$ and cathode rib width are also analyzed. The main conclusions are as follows:

(1) There is a strong correlation between the concentration of oxygen and $h_{c,GDL}$ in the CL. As $h_{c,GDL}$ decreases, the path of water removal and reactant gas transfer is shortened, enhancing the capabilities of water removal and reactant gas transfer in the porous medium. Thus, the cell performance of the CFRFF design improves as $h_{c,GDL}$ decreases.

(2) There is an optimal $h_{c,GDL}$ that results in optimal cell performance for the CRFF design. Because of the accumulation of under-rib liquid water, the optimal value of $h_{c,GDL}$ is 130 μm in this study.

(3) At a thin $h_{c,GDL}$, for the CRFF fuel cell, as the rib width increases, the under-rib oxygen transport region becomes larger, and the concentration polarization loss increases. For the CFRFF fuel cell, the increase in rib width can weaken the capability of water removal, increase the content of dissolved water in the ionomer, and reduce the Ohmic polarization loss. Therefore, under the condition of thin GDLs, the rib width of the CRFF fuel cell should be designed as small as possible, while the rib width of the CFRFF fuel cell can be slightly larger.

(4) The decrease in $h_{a,GDL}$ of the CRFF and CFRFF fuel cells reduces the Ohmic polarization loss by increasing the content of dissolved water in the ionomer, thus slightly improving the performance of the fuel cell and having little effect on the concentration polarization loss.

## Acknowledgements

This work is supported by the National Natural Science Foundation of China (Grant Nos. 51920105010 and 51921004) and the Department of Science and Technology of Inner Mongolia Autonomous Region (Grant No. 2022JBGS0027).